\documentclass[aps,11pt,nofootinbib,groupedaddress,superscriptaddress,preprintnumbers,showpacs]{revtex4}

\usepackage{amssymb, amsfonts, amsmath, bm}
\usepackage
{graphicx,color}
\usepackage{mathrsfs}

\begin{document}
\title{
{\Large
Self-similar motion of a Nambu-Goto string
}}

\hfill{\small RUP-16-24, KOBE-COSMO-16-08}

\author{Takahisa Igata}
\email{igata@rikkyo.ac.jp}
\affiliation{Department of Physics, Rikkyo University, Toshima, Tokyo 175-8501, Japan}
\author{Tsuyoshi Houri}
\email{houri@phys.sci.kobe-u.ac.jp}
\affiliation{Department of Physics, Kobe University, Kobe, Hyogo 657-8501, Japan}
\author{Tomohiro Harada}
\email{harada@rikkyo.ac.jp}
\affiliation{Department of Physics, Rikkyo University, Toshima, Tokyo 175-8501, Japan}

\pacs{11.25.-w, 11.27.+d, 98.80.Cq}

\begin{abstract}
We study the self-similar motion of a string 
in a self-similar spacetime by 
introducing the concept of a self-similar string, 
which is defined as the world sheet to which 
a homothetic vector field is tangent. 
It is shown that in Nambu-Goto theory, 
the equations of motion for a self-similar string 
reduce to those for a particle. 
Moreover, under certain conditions such as 
the hypersurface orthogonality of the homothetic vector field, 
the equations of motion for a self-similar string 
simplify to the geodesic equations on a (pseudo) 
Riemannian space. 
As a concrete example, 
we investigate a self-similar Nambu-Goto string 
in a spatially flat 
Friedmann-Lema\^itre-Robertson-Walker expanding universe with self-similarity 
and obtain solutions of open and closed strings, 
which have various nontrivial configurations 
depending on the rate of the cosmic expansion. 
For instance, we obtain a circular solution that 
evolves linearly in the cosmic time 
while keeping its configuration by the balance between the effects of 
the cosmic expansion and string tension. 
We also show the instability for linear radial perturbation of the circular solutions. 
\end{abstract}
\maketitle

\section{Introduction}
\label{sec:1}

In the context of modern physics, 
a string has been of strong interest for a long time. 
At large scales, 
one-dimensional topological defects formed in the early Universe 
are referred to as cosmic strings, 
which have sizes as large as each cosmological horizon~\cite{Vilenkin:2000jqa}. 
At small scales, 
strings are thought of as elementary components 
in string theories~\cite{Zwiebach:2004tj}. 
Under these circumstances, 
it is important to develop our understanding of classical string dynamics 
in curved spacetimes.

In Nambu-Goto string theory, 
string dynamics is governed by 
second-order nonlinear partial differential equations in two variables. 
Although some solutions are not always analytically tractable, 
some string solutions have been constructed 
with the aid of \textit{symmetry} 
in the following 
target spacetimes: 
Minkowski~\cite{Ogawa:2008qn, Burden:1982zb, Burden:1984xk, Embacher:1992te, Kozaki:2009jj, Igata:2012kx}，
black hole~\cite{Carter:1989bs, deVega:1996mv, Frolov:1996xw, Igata:2009dr, Igata:2009fd}, 
and cosmological spacetimes~\cite{Vilenkin:1981kz, Gasperini:1990xg, deVega:1993rm, Larsen:1995vr, Li:1993qc}. 
For example, 
stationarity of a string is defined by 
a timelike Killing vector field in a target spacetime, 
which is tangent to the world sheet. 
Since the Killing vector field generates the time evolution 
of the string, the equation of motion 
only determines its configuration on a time slice. 
This idea was formulated as 
\textit{the stationary strings}~\cite{Frolov:1988zn}, 
for which the Nambu-Goto equation
reduces to a geodesic equation, 
which is ordinary differential equations in a single variable. 
It was generalized to \textit{the cohomogeneity-one strings}~\cite{Ishihara:2005nu}, 
which are defined as the world sheet 
to which any Killing vector field (not necessarily stationary) is tangent. 
In physically realistic systems, 
these are candidates for final states
after radiating gravitational waves~\cite{Ogawa:2008yx}.

The Killing vector field can be generalized 
to the homothetic vector field, 
which is associated with self-similarity of a spacetime. 
A spacetime that admits a homothetic vector field is called 
a self-similar spacetime. 
This has been widely studied 
\cite{Harada:2001nh, Maeda:2001jh, Harada:2001hk, Maeda:2002br, Maeda:2002bv, Harada:2002ui, 
Maeda:2003vc, Harada:2003jg, Maeda:2004kw, Harada:2006dv, Harada:2007tj, Maeda:2007tk, Harada:2008rx} 
and was highlighted as the self-similar hypothesis~\cite{Carr:2005uf}, 
which states that 
solutions in general relativity naturally develop 
toward a self-similar form asymptotically 
under some physical circumstances. 
It, therefore, 
may be reasonable to consider 
self-similar spacetimes to be physically realistic 
in cosmology and astrophysics.

We are now in the position of having an interest in the dynamics of 
a string with self-similarity on a self-similar spacetime, that is, 
a self-similar string. 
Since this is a generalization of 
a stationary string on a stationary spacetime, 
we can expect a self-similar string to be a candidate for a final state 
in a self-similar spacetime. 
These solutions will model cosmic strings in the late time of an expanding universe. 
The purpose of this paper therefore is 
to formulate a self-similar string and 
to demonstrate its qualitative behavior.

This paper is organized as follows. 
In the following section, 
we propose a definition of a self-similar string 
in terms of self-similarity on a self-similar target spacetime. 
On the basis of Nambu-Goto string theory, 
we obtain the equation of motion for 
a self-similar string. 
In Sec.~\ref{sec:3}, 
we apply our formalism to a self-similar Nambu-Goto string 
in a spatially flat Friedmann-Lema\^itre-Robertson-Walker(FLRW)
expanding universe with 
self-similarity, which 
includes the Minkowski spacetime as a special case. 
Solving it in some integrable cases, 
we consider its dynamics through analytical models.  
Section \ref{sec:4} is devoted to a summary. 
Throughout this paper, we use geometrized units, in which $G=1$ and $c=1$.

\section{Formulation of a self-similar string}
\label{sec:2}

Let $(M,g_{\mu\nu})$ be a $D$-dimensional self-similar spacetime, 
i.e., a spacetime
that admits a homothetic vector field $\xi^\mu$, which is defined by
\begin{align}
\label{eq:HVeq}
\pounds_\xi g_{\mu\nu}=2\,C\,g_{\mu\nu},
\end{align}
where $\mu$ and $\nu$ have the range $0, 1, \ldots, D-1$, 
the left-hand side denotes 
the Lie derivative of $g_{\mu\nu}$ 
with respect to $\xi^\mu$, 
and $C$ is a constant. 
A homothetic vector field is said to be proper 
if $C\neq 0$, which is unique up to 
a Killing vector field. 
In the case where $C=0$, 
it is nothing but a Killing vector field. 
Given a homothetic vector field $\xi^\mu$, 
we are able to introduce a local coordinate system 
$x^\mu=(\eta, \mbox{\boldmath $x$})$ on $M$
such that 
\begin{align}
\label{eq:partialeta}
\xi^\mu \partial_\mu=\partial_\eta, 
\end{align}
where $\eta$ is called a homothetic coordinate 
and $\mbox{\boldmath $x$}=(x^1,x^2,\ldots, x^{D-1})$. 
In this coordinate system, 
if $\xi^\mu$ is timelike or spacelike, 
$g_{\mu\nu}$ is 
generally written 
in the $1+(D-1)$ form
\begin{align}
\label{eq:metric}
&ds^2=\Omega^2(\eta)\left[\,
f(\mbox{\boldmath $x$})
\left(d\eta+B_i(\mbox{\boldmath $x$})\,dx^i\right)^2
+h_{ij}(\mbox{\boldmath $x$})\,dx^i\,dx^j
\,\right],
\end{align}
where the Latin indices $i$, $j$, $\ldots$ have the range $1,2, \ldots, D-1$. 
Note that the functions $f$, $B_i$, and $h_{ij}$ 
do not depend on $\eta$, and 
$\Omega(\eta)=e^{C \eta}$ by virtue of Eq.~\eqref{eq:HVeq}. 
For a null $\xi^\mu$, the form (3) does not apply.\footnote{
We can construct a self-similar null string in the same way as in Ref.~\cite{Schild:1976vq} 
because a null $\xi^\mu$ is tangent to a null geodesic. 
}

Now let us define a self-similar string 
in terms of self-similarity of $(M, g_{\mu\nu})$ by 
employing the way that 
was introduced to define a cohomogeneity-one string~\cite{Frolov:1988zn, Ishihara:2005nu}. 
Let $X^\mu(\tau, \sigma)$ be embedding functions of a string, 
which describes a two-dimensional world sheet. 
Then, we define a self-similar string as a world sheet $\Sigma$ to which 
a homothetic vector field $\xi^\mu$ is tangent, that is, 
\begin{align}
\label{eq:self-similarity}
\xi^{[\,\mu}\dot{X}^{\nu}X'^{\lambda\,]}=0,
\end{align}
where $\lambda=0,1,\ldots, D-1$, 
the square bracket denotes antisymmetrization, 
and the dot and prime are
derivatives with respect to 
$\tau$ and $\sigma$, respectively. 
Since a homothetic vector field $\xi^\mu$ is 
tangent to $\Sigma$, 
the homothetic coordinate $\eta$ associated with $\xi^\mu$ can 
be introduced as a coordinate on $\Sigma$. 
Therefore, we may parametrize 
the embedding functions
in the homothetic coordinate system $(\eta, \mbox{\boldmath $x$})$ as 
\begin{align}
\label{eq:SSS}
X^\mu(\tau,\sigma)
=\left(\tau, \mbox{\boldmath $X$}(\sigma)\right),
\end{align}
where 
$\mbox{\boldmath $X$}(\sigma)=(X^1(\sigma),\ldots, X^{D-1}(\sigma))$. 
This expression indeed satisfies Eq.~\eqref{eq:self-similarity}. 
Once a theory is specified, 
the equations of motion for $\mbox{\boldmath $X$}(\sigma)$ reduce to ordinary 
differential equations. 
Thus, the problem of determining the dynamics of a self-similar string 
reduces to that of ``a particle." 
However, it is still uncertain whether 
all the equations are compatible because 
the equations for $\mbox{\boldmath $X$}(\sigma)$ can be 
overdetermined. 
In what follows, 
we restrict ourselves to consider a self-similar string 
in Nambu-Goto theory. 
Then, we see that 
its equations of motion result in those of a particle motion.

\subsection{Nambu-Goto equation for a self-similar string}
\label{sec:2A}

Let us assume that 
the dynamics of a self-similar string on $(M, g_{\mu\nu})$ is governed by 
Nambu-Goto string theory. 
The Nambu-Goto action is given by
\begin{align}
S=-\mu \int_\Sigma \sqrt{-\gamma}\,d\tau\,d\sigma,
\label{eq:action}
\end{align}
where $\mu$ is the string tension, 
and $\gamma$ is defined as the determinant of the induced metric 
$\gamma_{ab}=g_{\mu\nu}(X)\,\partial_a X^\mu \partial_b X^\nu$ on $\Sigma$, 
where $a$, $b$ have $\tau$, $\sigma$. 
When no confusion arises from abbreviation, 
we omit the arguments $X^\mu$ from any function 
in what follows. 
The variation of Eq.~\eqref{eq:action} 
with respect to $X^\mu$ yields 
the Nambu-Goto equation
\begin{align}
\label{eq:N-Geq}
\partial_a\left(\sqrt{-\gamma}\,\gamma^{ab}\partial_b X^\mu\right)
+\sqrt{-\gamma}\,\gamma^{ab}\,\Gamma^\mu{}_{\nu\lambda}\,
\partial_aX^\nu \partial_bX^\lambda=0,
\end{align}
where $\gamma^{ab}$ is the inverse metric of $\gamma_{ab}$, 
and $\Gamma^\mu{}_{\nu\lambda}$ denotes the Christoffel symbol of $g_{\mu\nu}$. 
Under the settings of Eqs.~\eqref{eq:metric}--\eqref{eq:SSS},
the induced metric has the form
\begin{align}
\label{eq:gamma_ab}
\gamma_{ab}\,d\zeta^a d\zeta^b=\Omega^2
\left[\,
f\left(d\tau+B\,d\sigma\right)^2+h\,d\sigma^2
\,\right],
\end{align}
where 
$\zeta^a=(\tau, \sigma)$ 
and
\begin{align}
&B=B_i\,X'^i,
\\
&h=h_{ij}\,X'^iX'^j. 
\end{align}
The determinant of $\gamma_{ab}$ is given by $\gamma=\Omega^4f h$. 
Then the components of $\sqrt{-\gamma}\,\gamma^{ab}$ 
in terms of $\zeta^a$ are calculated as
\begin{align}
\label{eq:gamma2}
\sqrt{-\gamma}\,\gamma^{ab}=-\mathrm{sgn}(f)\,\sqrt{-f/h}\left(
\begin{array}{cc}
B^2+h/f&-B\\
-B&1
\end{array}
\right).
\end{align}
Substituting Eq.~\eqref{eq:gamma2} into Eq.~\eqref{eq:N-Geq} and 
dividing it by $\sqrt{-\gamma}\,\gamma^{\sigma\sigma}$, 
we obtain $E^\mu=0$, where 
\begin{align}
E^\mu
=&\ 
X'^\nu \nabla_\nu X'^\mu
+
\left(\ln\sqrt{-f/h}\right)'
\,X'^\mu
\cr
&
+\left(B^2+h/f\right)
\dot{X}^\nu \nabla_\nu \dot{X}^\mu
-\frac{(B\,\sqrt{-f/h})'}{\sqrt{-f/h}}\,\dot{X}^\mu
-2B\,\Gamma^\mu{}_{\nu\lambda}\dot{X}^\nu X'^\lambda, 
\end{align}
and $\nabla_\mu$ is the Levi-Civita covariant derivative associated with $g_{\mu\nu}$. 
Each component of $E^\mu=0$ has the form
\begin{align}
\label{eq:N-Geq1}
&E^\eta=-B_i\,E^i=0,
\\
\label{eq:N-Geq2}
&E^i=
\mathcal{E}^i
+2\,C\left(
B^ih_{jk}-\delta^i{}_jB_k 
\right)X'^j X'^k=0,
\end{align}
where $B^i=h^{ij}B_j$, $h^{ij}$ is the inverse of $h_{ij}$, 
and we have defined $\mathcal{E}^i$ as
\begin{align}
\label{eq:geodesic}
\mathcal{E}^i= 
X'^jD_j X'^i
-\frac{d\ln \sqrt{-fh}}{d\sigma}\,X'^i,
\end{align}
where $D_i$ denotes the Levi-Civita 
covariant derivative associated with $|\,f\,|\,h_{ij}$. 
It immediately follows that a solution of Eq.~\eqref{eq:N-Geq2} 
always solves Eq.~\eqref{eq:N-Geq1}.  
This shows that the assumptions for a self-similar string 
are compatible with Nambu-Goto theory. 
Thus, we have obtained the reduced Nambu-Goto equation \eqref{eq:N-Geq2}
for a self-similar string in a self-similar spacetime, 
which is identical to the equations of motion 
for a particle in $D-1$ dimensions, as was expected.

In Refs.~\cite{Frolov:1988zn, Ishihara:2005nu}, 
the equations of motion for a cohomogeneity-one string were obtained 
by just taking the variation of the reduced action. 
However, the present analysis does not rely on this method. In fact, the 
reduced action for
a self-similar string, 
along with the assumptions \eqref{eq:partialeta}--\eqref{eq:SSS}, 
is given by
\begin{align}
\label{eq:reduced action}
S=-\mu \int \Omega^2 \sqrt{-f\,h}\,d\tau\,d\sigma. 
\end{align}
The variation of this reduced action leads to ${\cal E}^i=0$, 
which is not equivalent to the 
the equations of motion~\eqref{eq:N-Geq2} in general. 
This difference commonly happens 
when we consider a reduced action, 
depending on how compatible the assumptions we consider are. 
Our calculation shows that 
if the second term in Eq.~\eqref{eq:N-Geq2} vanishes 
the equations obtained by taking the variation of 
the reduced action~\eqref{eq:reduced action} 
coincide with the equations of motion for a self-similar string. 
In the next section, 
we investigate in detail the conditions that 
the second term in Eq.~\eqref{eq:N-Geq2} vanishes.

\subsection{Reduction to a geodesic equation}
\label{sec:2B}

In particular situations where
\begin{align}
\label{eq:cond}
C\left(
B^ih_{jk}-\delta^i{}_jB_k 
\right)X'^j X'^k=0,
\end{align}
the reduced Nambu-Goto equation \eqref{eq:N-Geq2} further 
simplifies to $\mathcal{E}^i=0$, i.e., 
\begin{align}
\label{eq:geod}
X'^jD_j X'^i
=\frac{d\ln \sqrt{-fh}}{d\sigma}\,X'^i.
\end{align}
This equation describes a geodesic flow on the 
orbit space $O$ of 
$\xi^\mu$ 
with the metric $|\,f\,|\,h_{ij}$. 
Hence, 
the problem of finding a self-similar Nambu-Goto string 
on $(M, g_{\mu\nu})$ is simplified to that of solving 
the geodesic equation 
on a (pseudo) Riemannian space $(O, |\,f\,|\,h_{ij})$.

This situation can occur if any one of the conditions
\begin{align}
\label{eq:3cases}
\begin{tabular}{ll}
\textrm{(i)}&
~$C=0$,
\\
\textrm{(ii)}&
~$B_i=0$,
\\
\textrm{(iii)}&
~$B^i \parallel X'^i$
\end{tabular}
\end{align}
is satisfied. 
Since Condition (i) corresponds to the case of a cohomogeneity-one string, 
our formulation justifies 
the reduced action method by use of 
a Killing vector field 
developed in Refs.~\cite{Frolov:1988zn, Ishihara:2005nu}. 
Condition (ii) means that 
the homothetic vector field is hypersurface orthogonal 
and 
certainly occurs as is seen in Sec.~\ref{sec:3}. 
Condition (iii) can be rewritten as 
$\xi_i  \parallel  h_{ij}\,X'^j$.

\section{Self-similar Nambu-Goto string}
\label{sec:3}

\subsection{Formulation in a self-similar expanding universe}
\label{sec:3a}
In this section, 
we consider the dynamics 
of a self-similar Nambu-Goto string in an 
expanding flat universe described by the 
FLRW metric 
\begin{align}
\label{eq:RW}
ds^2=-dt^2+a^2(t)\left(dx^2+dy^2+dz^2\right), 
\end{align}
where $(t,x,y,z)$ are the comoving Cartesian coordinates, and $a(t)$ is the scale factor. 
For the metric to admit a proper homothetic vector field $\xi^\mu$, 
the scale factor is restricted to have the form~\cite{Maartens:1986}
\begin{align}
\label{eq:a}
a(t)=\left(t/t_0\right)^{1-1/C},
\end{align}
where $C$ and $t_0$ are nonzero constants, and then $\xi^\mu$ 
is given by
\begin{align}
\label{eq:HVRW}
\xi^\mu \partial_\mu
=
C \,t\, \partial_t+r\,\partial_r. 
\end{align}
The metric coincides with the 
Minkowski one when $C=1$. 
For $C\neq1$, we take the sign of $t_0$ to be $t/t_0>0$. 
Furthermore, we assume that a universe is expanding,
$da/dt>0$. 
Thus, under the choice of $\partial_t$ to be future directed, 
we take the coordinate range $0<t<\infty$ for $C>1$ or $C<0$ 
and $-\infty<t<0$ for $0<C<1$.  
If we consider that 
a universe is filled with a perfect fluid with 
the equation of state $p=w\,\rho$, 
the parameters $C$ and $w$ are related to each other as
\begin{align}
w=\frac{3-C}{3\left(C-1\right)}, 
\end{align}
where $w$ is a constant other than $-1$ and $-1/3$.

Moreover, we introduce a conformal time $\lambda$ defined by
\begin{align}
\lambda
=\lambda_0\left(t/t_0\right)^{1/C},
\end{align}
where $\lambda_0=C\,t_0$. 
The metric then takes the form
\begin{align}
\label{eq:cflatmetric}
ds^2
&=\left(\lambda/\lambda_0
\right)^{2\left(C-1\right)}
\left(-d\lambda^2+dr^2+r^2\,d\Omega^2\right),
\end{align}
where we have introduced the spherical-polar coordinates $(r,\theta,\phi)$ in 
the spatial part and 
$d\Omega^2=d\theta^2+\sin^2\theta\,d\phi^2$. 
The homothetic vector field is given by 
$\xi^\mu \partial_\mu=\lambda\,\partial_\lambda+r\,\partial_r$. 
In the following subsections, 
we introduce a local homothetic coordinate system in Region I ($r<|\,\lambda\,|$) 
and Region I\hspace{-.1em}I ($|\,\lambda\,|<r$), where 
the homothetic vector field is timelike and spacelike, respectively, 
and derive 
the equations of motion for a self-similar  
string associated with $\xi^\mu$.
Hereafter, we adopt the units in which $|\,\lambda_0\,|=1$.

\subsubsection{Region I (\,$r<|\,\lambda\,|$)}
\label{sec:3A1}

Let us make the coordinate transformation in Region I defined by
\begin{align}
\label{eq:lambda}
&\lambda=\kappa \,e^{\eta}\,\cosh \chi,
\\
\label{eq:r}
&r=e^{\eta}\,\sinh \chi,
\end{align}
where $\kappa=\mathrm{sgn}(\lambda_0)$, and
the coordinates $\eta$ and $\chi$ have the range 
$-\infty < \eta<\infty$ and $0<\chi<\infty$, respectively．
The metric is then transformed to
\begin{align}
\label{eq:RWmetric}
ds^2&=e^{2C \eta}\left(\cosh \chi\right)^{2\left(C-1\right)}\left(-d\eta^2+d\chi^2+\sinh^2\chi\,d\Omega^2\right),
\end{align}
which in particular yields the Milne metric in $C=1$.

Since the form \eqref{eq:RWmetric} fits into Eq.~\eqref{eq:metric}, 
the homothetic vector field \eqref{eq:HVRW} has the same form as Eq.~\eqref{eq:partialeta}. 
We also notice that the metric satisfies Condition (ii) discussed in Sec.~\ref{sec:2B}, 
so that the dynamics of a self-similar string is 
determined by the geodesic equation~\eqref{eq:geod}. 
In the present homothetic coordinate system 
$(\eta, \chi, \theta, \phi)$, the embedding functions are written as 
\begin{align}
X^\mu(\tau,\sigma)=
\left(\tau, \mathcal{X}(\sigma), \Theta(\sigma), \Phi(\sigma)\right),
\end{align}
where $X^i(\sigma)$ are determined by the geodesic equation 
with respect to the metric 
\begin{align}
|\,f\,|\,h_{ij}\,dx^i\,dx^j
=
\left(\cosh \chi\right)^{4\left(C-1\right)}\left(
d\chi^2+\sinh^2\chi\,d\Omega^2\right),
\end{align}
which is conformal to the three-dimensional hyperbolic space
$\mathbb{H}^3$. 
In particular, a solution $X^i(\sigma)$ for $C=1$ is identified with a geodesic on $\mathbb{H}^3$.

Let us solve the geodesic equation by using the Hamiltonian formalism. 
To rewrite this system in canonical variables, 
we employ the Polyakov-type action with the Lagrangian in the form
\begin{align}
\label{eq:L}
L=
\frac{1}{2N}\,|\,f\,|\,h_{ij}\frac{dX^i}{d\sigma}\frac{dX^j}{d\sigma}+\frac{N}{2},
\end{align}
where $N$ is an auxiliary variable. 
Defining the canonical momentum 
$p_i=N^{-1}|\,f\,|\,h_{ij}X'^{j}$ 
conjugate to $X^i$, 
we obtain the Hamiltonian
\begin{align}
&H=\frac{N}{2}\left(
\frac{p_\chi^2}{\left(
\cosh \chi
\right)^{4\left(C-1\right)}}
+\frac{l^2}{\left(
\cosh \chi
\right)^{4\left(C-1\right)}\sinh^2\chi}-1
\right),
\end{align}
where, without loss of generality, 
we have assumed $\Theta(\sigma)=\pi/2$, $p_\theta(\sigma)=0$, 
and $p_\phi=l\geq0$, 
because spherical symmetry is induced on  $(O, |\,f\,|\,h_{ij})$. 
Since this Hamiltonian does not depend on $\phi$, 
the quantity $l$ is a constant of motion and 
is related to the strength of the conserved angular momentum flux 
on the world sheet.

The Hamilton equation and the constraint $H=0$ yield
\begin{align}
\label{eq:Chi'}
&\mathcal{X}'^2+V(\mbox{\boldmath $X$})=0,
\\
\label{eq:V}
&V=\frac{l^2}{\sinh^2\chi}-\left(
\cosh \chi
\right)^{4\left(C-1\right)},
\\
\label{eq:Phi'}
&\Phi'=\frac{l}{\sinh^2\mathcal{X}},
\end{align}
for which the gauge is fixed by 
$N=\left(\cosh\mathcal{X}\right)^{4(C-1)}$. 
Equations~\eqref{eq:Chi'} and \eqref{eq:V} give us
a one-dimensional problem with the potential $V$. 
The first term of $V$ is related to the angular momentum flux 
and makes a potential barrier near $\chi=0$. 
The second term represents the effect of the cosmic expansion 
because this includes $C$ and turns to be flat in the Minkowski background, $C=1$.

Let us classify the behavior of solutions in terms of $l$ and $C$. 
Since the $C=1$ case is analyzed in detail in Sec.~\ref{sec:3B1}, 
we examine the case $C\neq 1$ in what follows.

\begin{figure}[t]
\centering
\includegraphics[width=14.8cm,clip]{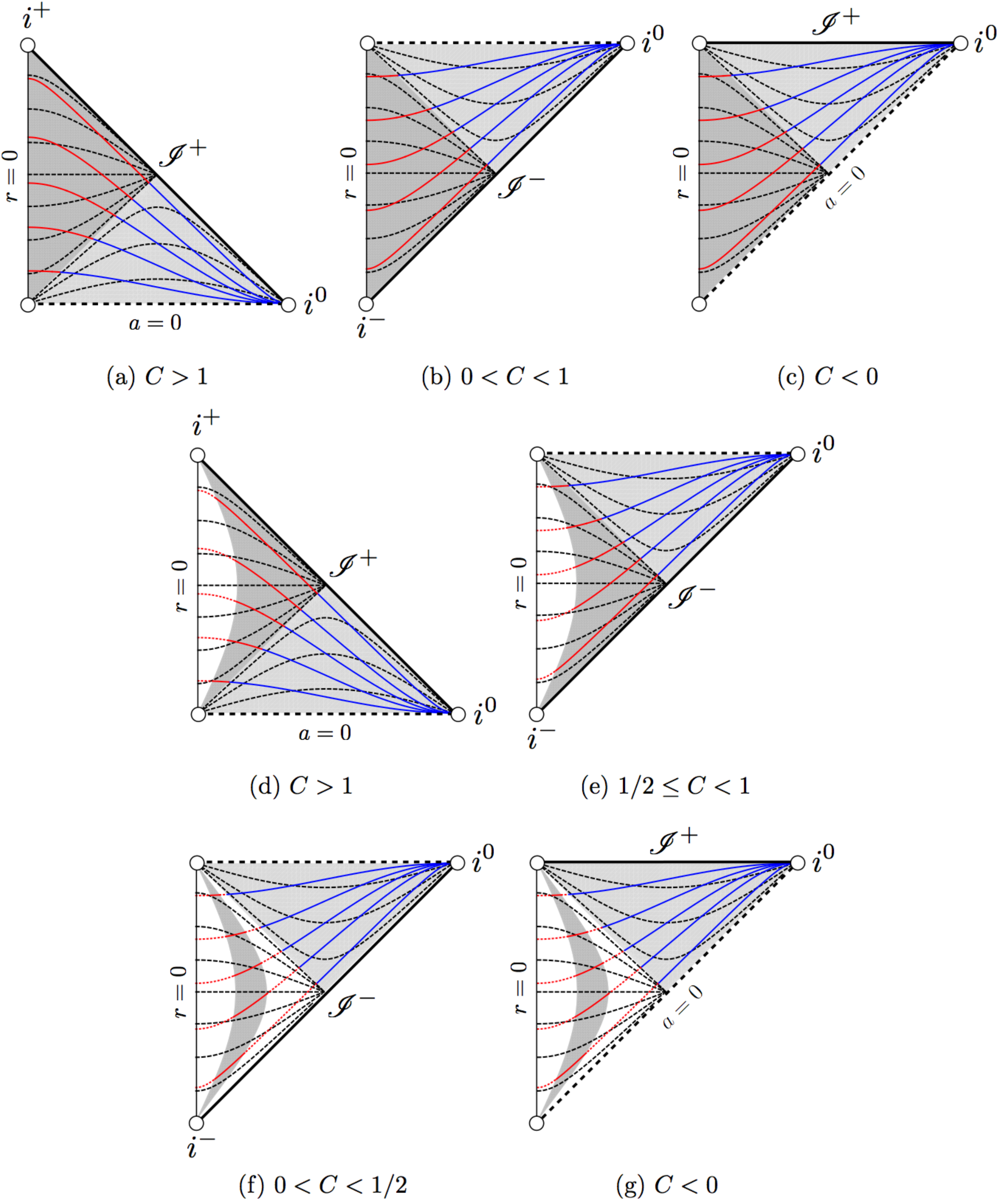}
 \caption{
Embeddings of a self-similar Nambu-Goto string into 
the conformal diagram of the spatially flat 
FLRW
spacetime with self-similarity. 
Figures~(a)--(c) are the case 
$l=0$ and $\bar{l}=0$, and Figs.~(d)--(g) are the case 
$l\neq0$ and $\bar{l}\neq0$, 
in which $l$ is restricted to $0< l<l_\textrm{c}$ for 
$C\leq 1/2$. 
The symbols 
$i^0$, $i^{\pm}$, and $\mathscr{I}^{\pm}$ are 
spatial, timelike, and null infinity, 
respectively, 
$+$ and $-$ of which indicate future and past, respectively. 
The black thick dashed line is a singularity. 
The dark and light gray regions show
a world sheet in Region I and in Region I\hspace{-.1em}I, respectively. 
The black thin dashed lines denote 
the constant $\eta$ slices in Region I and 
the constant $\chi$ slices in Region I\hspace{-.1em}I. 
The red and blue solid lines are the constant $t$ slices. 
}
 \label{fig:1}
\end{figure}

We consider qualitative properties of a solution with $l=0$. 
Since $l=0$ leads to $V<0$ and $\Phi'=0$, 
we obtain a straight string on each of the constant $\eta$ slices
that 
passes through $\chi=0$ and asymptotically approaches $\chi=\infty$. 
On the other hand, 
the configuration on each of the constant $t$ slices
is a finite straight segment 
with the end points moving at the speed of light. 
We can explicitly see the dependence of string configuration on time slices in 
Figs.~\ref{fig:1}(a)--\ref{fig:1}(c), 
which show these embeddings into the conformal diagrams of 
the spatially flat FLRW spacetimes.  
The dark gray region denotes the world sheet, 
and black dashed and red solid lines in Region I 
denote the constant $\eta$ and constant $t$ slices, 
respectively. 
Note that this solution is eventually identified with 
a cohomogeneity-one string that possesses spatial homogeneity, 
and in addition,  
its analytic continuation to Region I\hspace{-.1em}I
provides a straight line with infinite length passing through $r=0$ on the 
constant $t$ slices. 
As shown in Ref.~\cite{Vilenkin:1981kz}, 
a straight line solution is stable in linear perturbations. 
To focus on a self-similar string, we assume $l>0$ in what follows.

For $C>1/2$ (\,i.e., $w<-5/3$ or $w>-1/3$),
$V$ always has 
a zero at $\chi=\chi_*$, 
where $V(\chi_*)=0$ and is negative for all $\chi\geq \chi_*$. 
It follows that 
a solution on 
each of the constant $\eta$ slices extends from $\chi=\chi_*$ to $\chi=\infty$ (null infinity). 
In Figs.~\ref{fig:1}(d) and \ref{fig:1}(e), 
the world sheets are depicted with dark gray in $\chi\geq\chi_*$ of Region I.

For $C<1/2$ (\,i.e., $-5/3<w<-1/3$\,), 
$V$ has at most two zeros, 
$\chi_\textrm{min}$ and $\chi_\textrm{max}$. 
The necessary and sufficient condition for $V$ to have zeros is 
\begin{align}
\label{eq:lrange}
&0<  l \leq l_\textrm{c}
=\frac{\left(1-2\,C\right)^{(1-2\,C)/2}}{\left[\,2\left(1-C\right)\,\right]^{1-C}},
\end{align}
where the equality $l=l_\textrm{c}$ is achieved if 
$\mathcal{\chi}_\textrm{min}=\mathcal{\chi}_\textrm{max}$. 
The allowed range of a solution satisfying Eq.~\eqref{eq:lrange}
 is restricted to 
$\chi_\textrm{min}\leq \chi \leq \chi_\textrm{max}$, 
which implies that a closed string configuration can be included. 
Figures~\ref{fig:1}(f) and \ref{fig:1}(g) show 
a string extended over a finite spatial section in Region I. 
The special case $\mathcal{\chi}_\textrm{min}=\mathcal{\chi}_\textrm{max}$ 
is analyzed in detail in Sec.~\ref{sec:3B3}.

The remaining is $C=1/2$ (\,i.e., $w=-5/3$), 
of which the FLRW universe is filled with 
phantom energy with special properties~(see, for example, \cite{Dabrowski:2003jm}).
Because $V$ can be negative for all $\chi\geq \tanh^{-1}l$, where $0<l<1$, 
a solution on the constant $\eta$ slices extends from $\chi= \tanh^{-1}l$ to $\chi=\infty$ 
(past null infinity), 
as can be seen in Fig.~\ref{fig:1}(e). 
We are able to obtain an analytical solution in Sec.~\ref{sec:3B2}.

\medskip

The end of this section is devoted to reconsidering 
the dynamical evolution of a self-similar string in Region I  
in the comoving coordinate system. 
If we synchronize the string time coordinate to the comoving time $t$, 
then in the comoving spherical-polar coordinate system  
a radial solution in the proper length is written as 
$a(t)\, r(t,\sigma) =\kappa\,t\,\tanh \mathcal{X} (\sigma)
$. 
Hence, in the comoving Cartesian coordinate system, we have
\begin{align}
a(t)\, X^i(t, \sigma) = t\,Q^i(\sigma),
\end{align}
where $Q^i(\sigma)$ are functions of 
$\mathcal{X}(\sigma)$, $\Theta(\sigma)$, and $\Phi(\sigma)$. 
This means the homothetic scaling of a self-similar string configuration 
as $t$ proceeds.

\subsubsection{Region I\hspace{-.1em}I (\,$|\,\lambda\,|<r$)}
\label{sec:3A2}

Let us turn our attention to a self-similar Nambu-Goto string 
in Region I\hspace{-.1em}I. 
We define new coordinates $\bar{\chi}$ and $\bar{\eta}$ as 
\begin{align}
&\lambda=\kappa\,e^{\bar{\eta}}\sinh \bar{\chi},
\\
&r=e^{\bar{\eta}}\cosh \bar{\chi},
\end{align}
where 
$0< \bar{\chi} < \infty$ and 
$-\infty<\bar{\eta}<\infty$, 
so that Eq.~\eqref{eq:cflatmetric} 
is transformed as 
\begin{align}
\label{eq:RWmetric2}
ds^2=e^{2C\bar{\eta}}\left(\sinh \bar{\chi}\right)^{2\left(C-1\right)}\left(-d\bar{\chi}^2+d\bar{\eta}^2+\cosh^2\bar{\chi}\,d\bar{\Omega}^2\right). 
\end{align}

Since this form of the metric~\eqref{eq:RWmetric2} fits into Eq.~\eqref{eq:metric} 
and also
satisfies Condition (ii) in Sec.~\ref{sec:2B}, 
the dynamics of a self-similar string associated with the 
homothetic vector field \eqref{eq:HVRW} 
is determined by the geodesic equation~\eqref{eq:geod}. 
In this coordinate system, $\left(\bar{\eta}, \bar{\chi}, \bar{\theta}, \bar{\phi}\right)$, 
the embedding functions are written as
\begin{align}
X^\mu(\tau,\sigma)=\left(\sigma, \bar{\mathcal{X}}(\tau), \bar{\Theta}(\tau), \bar{\Phi}(\tau)\right),
\end{align}
where we have interchanged $\tau$ and $\sigma$ 
in comparison to Eq.~\eqref{eq:SSS} 
because $\xi^\mu$
is spacelike in Region I\hspace{-.1em}I. 
The embedding functions $\bar{X}^i(\sigma)$ are determined 
by the geodesic equation with respect to the metric
\begin{align}
|\,f\,|\,h_{ij}\,dx^i\,dx^j=\left(\sinh \bar{\chi}\right)^{2\left(C-1\right)}\left(
-d\bar{\chi}^2+\cosh^2\bar{\chi}\,d\bar{\Omega}^2\right).
\end{align}
In the same manner as used in Sec.~\ref{sec:3A1}, 
we obtain the Hamiltonian 
for a geodesic in $(O, |\,f\,|\,h_{ij})$
\begin{align}
&\bar{H}=\frac{\bar{N}}{2}\left(
-\frac{p_{\bar{\chi}}^2}{\left(
\sinh \bar{\chi}
\right)^{4\left(C-1\right)}}
+\frac{\bar{l}^{\,2}}{\left(
\sinh \bar{\chi}
\right)^{4\left(C-1\right)}\cosh^2\bar{\chi}}+1
\right),
\end{align}
where 
$\bar{N}$ is a Lagrange multiplier enforcing a constraint, and without 
loss of generality, 
we have assumed that $\bar{\Theta}(\tau)=\pi/2$, $p_{\bar{\theta}}(\sigma)=0$, and $\bar{l}=p_{\bar{\phi}}\geq0$. 
Then the Hamilton equation and the constraint $\bar{H}=0$ lead to
\begin{align}
\label{eq:dotChi}
&\dot{\bar{\mathcal{X}}}^2+\bar{V}(\mbox{\boldmath $X$})=0,
\\
\label{eq:Vbar}
&\bar{V}=-\frac{\bar{l}^{\,2}}{\cosh^2\bar{\chi}}-\left(\sinh \bar{\chi}\right)^{4\left(C-1\right)},
\\
\label{eq:dotPhi}
&\dot{\bar{\Phi}}=\frac{\bar{l}}{\cosh^2\bar{\mathcal{X}}},
\end{align}
where we have chosen 
$\bar{N}=\left(\sinh\bar{\mathcal{X}}\right)^{4(C-1)}$ in this gauge. 
Equations \eqref{eq:dotChi} and \eqref{eq:Vbar} 
give us a one-dimensional problem with the potential $\bar{V}$. 
Let us classify the behavior of the solutions 
in terms of $\bar{l}$ and $C$ in what follows.

For $C>1$ (i.e., $w>-1/3$), 
a self-similar string on each of the constant $\bar{\chi}$ slices has 
an end point at $\bar{\eta}=0$ on the initial singularity 
and asymptotically approaches $\bar{\eta}=\infty$ (spatial infinity).  
In the limit $\bar{\chi}\to 0$, 
this string reaches the initial singularity, 
and in the limit $\bar{\chi}\to \infty$, 
it approaches the null hypersurface to consist of 
$\lambda=r$ and part of future null infinity. 
In Figs.~\ref{fig:1}(a) and \ref{fig:1}(d), 
we can see the embedding of such solutions into the 
conformal diagrams of the spatially flat FLRW spacetimes, 
where the light gray region denotes the world sheet in Region I\hspace{-.1em}I.
The black thin dashed lines
denote the constant $\bar{\chi}$ slices, 
and the blue solid lines denote the constant $t$ slices.  
These figures explicitly show that 
a string on the constant $t$ slices in Region I\hspace{-.1em}I
has an end point at $r=|\,\lambda\,|$, 
which moves at the speed of light, 
and a boundary at spatial infinity.

For $0<C<1$ (i.e., $w<-1$), 
a self-similar string on the constant $\bar{\chi}$ slices
has an end point 
at $\bar{\eta}=0$ (big rip singularity), 
whereas the other asymptotically approaches $\bar{\eta}=\infty$ (spatial infinity), 
as seen in Figs.~\ref{fig:1}(b), \ref{fig:1}(e), and \ref{fig:1}(f). 
In the limit $\bar{\chi}\to \infty$, 
this string approaches the null hypersurface made up of 
$\lambda=-r$ and part of past null infinity and 
encounters the big rip singularity at $\bar{\chi}= 0$.

For $C<0$ (i.e., $-1<w<-1/3$), 
a solution on the constant $\bar{\chi}$ slices
shows that a string extends to 
two pieces of spatial infinity, 
as seen in Figs.~\ref{fig:1}(c) and \ref{fig:1}(g). 
In the limit $\bar{\chi}\to \infty$, 
this string reaches the null hypersurface composed of 
$\lambda=-r$ and part of the initial null singularity 
and asymptotically approaches $\bar{\chi}=0$ (future null infinity).

For $\bar{l}= 0$, 
a self-similar string 
results in a cohomogeneity-one string of half line 
with the end point moving at the speed of light. 
This analytical continuation to Region I 
can be an infinite straight line passing through $r=0$.  
We assume 
$\bar{l}>0$ in what follows.

\subsection{Analytical solutions}

In the following sections, 
we investigate analytical solutions to describe a self-similar string in 
an FLRW expanding universe with self-similarity. 

\subsubsection{$C=1$ (Minkowski spacetime)}
\label{sec:3B1}

Let us analyze a self-similar Nambu-Goto string associated with $\xi^\mu$ 
in $C=1$ (i.e., Minkowski spacetime). 
In Region I, we can obtain a solution from Eqs.~\eqref{eq:Chi'}--\eqref{eq:Phi'} in the form
\begin{align}
\label{eq:chi}
&\cosh\mathcal{X}=\sqrt{1+l^2}\,\cosh\sigma,
\\
\label{eq:cos1}
&\cos \Phi=\frac{l}{\sqrt{l^2+\tanh^2\sigma}},
\end{align}
where, without loss of generality, we have fixed each integral constant as 
$\cosh\mathcal{X}(0)=(1+l^2)^{1/2}$ and $\Phi(0)=0$ 
and have taken a branch in Eq.~\eqref{eq:cos1}. 
In the Cartesian coordinate system, this solution is rewritten in the form 
$X^\mu(\tau,\sigma)=(T(\tau,\sigma), X(\tau,\sigma), Y(\tau,\sigma), 0)$, 
where
\begin{align}
\label{eq:T}
&T
=
\kappa\,e^{\tau}\sqrt{1+l^2}\,\cosh \sigma,
\\
\label{eq:X}
&X=
l\,e^{\tau} \,\cosh \sigma,
\\
\label{eq:Y}
&Y=
e^{\tau} \,\sinh
\sigma. 
\end{align}
These functions indeed solve the two-dimensional wave equation 
$-\ddot{X}^{\mu}+X''^{\mu}=0$, 
which is derived from Eq.~\eqref{eq:N-Geq} 
because the gauge choice of this solution 
is the conformal gauge, 
$\sqrt{-\gamma}\,\gamma^{ab}=\eta^{ab}$.

As expected from the fact that Eqs.~\eqref{eq:chi} and \eqref{eq:cos1} 
are identical to a geodesic equation in $\mathbb{H}^3$, 
the solution describes a straight line on each of the 
constant $t$ slices such that 
\begin{align}
\label{eq:sol}
X(t,y)=\kappa\,\frac{l\,t}{\sqrt{1+l^2}},
\end{align}
where we have introduced $t$ and $y$ as new string parameters that 
must satisfy the inequality 
\begin{align}
\left|\,\frac{y}{t}\,\right|<\frac{1}{\sqrt{1+l^2}}.
\end{align}
This is required from the fact that $\xi^\mu$ is timelike in Region I.

Since the string moves in $x$-direction uniformly, 
we can always find the static frame of the string by applying 
the Lorentz transformation. 
Hence, without loss of generality, we may choose $l=0$. 
In the region $r<-t$, 
this solution shows a physical picture as follows: 
At an initial time, e.g., $t=-1$, 
the string is placed on the $y$ axis with the proper length $2$, 
shrinks to the length $2\,|\,t\,|$ at time $t$, 
which means that the end points approach each other at the speed of light, 
and finally collapses to a point in the limit $t \to 0$,  
whereas in the region $r<t$, we obtain its time reversal picture.

\medskip

Let us focus on a self-similar Nambu-Goto string in Region I\hspace{-.1em}I. 
We can solve Eqs.~\eqref{eq:dotChi}--\eqref{eq:dotPhi} as
\begin{align}
&\sinh \bar{\mathcal{X}}=\sqrt{1+\bar{l}^{\,2}}\,\sinh \tau,
\\
&
\label{eq:cos2}
\cos \bar{\Phi}=\frac{\bar{l}\,\tanh \tau}{\sqrt{1+\bar{l}^{\,2}\,\tanh^2\tau}},
\end{align}
where $0<\tau<\infty$, 
without loss of generality; 
we have chosen that $\sinh \bar{\mathcal{X}}(0)=0$ and $\cos \bar{\Phi}(0)=0$ 
and have taken a branch in Eq.~\eqref{eq:cos2}. 
In the Cartesian coordinates, this solution is written as 
$X^\mu(\tau,\sigma)=(\bar{T}(\tau,\sigma), \bar{X}(\tau,\sigma), \bar{Y}(\tau,\sigma), 0)$, 
where
\begin{align}
&\bar{T}=\kappa\,e^{\sigma}\,\sqrt{1+\bar{l}^{\,2}}\,
\sinh\tau,
\\
&\bar{X}=\bar{l}\,e^{\sigma}\,\sinh \tau,
\\
&\bar{Y}=e^{\sigma} \cosh\tau.
\end{align}
After the reparametrization to $t$ and $y$, 
the form of this solution coincides with Eq.~\eqref{eq:sol} 
under identifying $l$ to $\bar{l}$, 
where these string parameters are restricted by 
the condition that $\xi^\mu$
is spacelike, $\left|\,y/t\,\right|>1/(1+\bar{l}^{\,2})^{1/2}$. 
Hence, this solution describes a half line on each constant $t$ and 
moves in $x$-direction uniformly. 
The end point moves at the speed of light. 

These results explicitly show that 
the solution obtained in Region I 
can be analytically continued to Region I\hspace{-.1em}I 
through the boundary between these regions. 
The maximally extended string shows a straight line, 
which is cohomogeneity-one, with 
boundaries at spatial infinity.

\subsubsection{$C=1/2$}
\label{sec:3B2}

Let us consider a self-similar Nambu-Goto string associated with $\xi^\mu$
in the expanding universe with $C=1/2$, 
which is filled with the phantom energy of $w=-5/3$. 
Assuming $0<l<1$, in which Eq.~\eqref{eq:Chi'}
has a real solution as examined in Sec.~\ref{sec:3A1}, 
we obtain a solution in Region I
\begin{align}
\label{eq:Dcosh}
&\cosh \mathcal{X}=
\sqrt{(1-l^2)\,\sigma^2+\frac{1}{1-l^2}},
\\
&\tan \Phi=-\frac{l}{1-l^2}\,\sigma^{-1},
\end{align}
where, without loss of generality, we have chosen each integral constant as 
$\cosh \mathcal{X}(0)=1/(1-l^2)^{1/2}$ and $\Phi(0)=\pi/2$. 
Since $\mathcal{X}\to \infty$ as $\sigma \to \pm \infty$, 
this string on the constant $\eta$ slices 
possesses 
boundaries at $\chi=\infty$ (past null infinity) 
and, hence, is an open string with infinite length. 
In the comoving Cartesian coordinate system, 
on the other hand, 
the solution is of the form
\begin{align}
Y(t,x)=l\sqrt{(t/t_0)^4-x^2}, 
\end{align}
where $t_0=-2$ in our units, and 
we have introduced $t$ and $x$ as new string parameters. 
This shows a semiellipse on each 
constant $t$ 
and includes end points at $(x,y,z)=(\pm(t/t_0)^2,0 ,0)$, 
where each segment moves at the speed of light. 
These two points of view are 
illustrated in Fig.~\ref{fig:1}(e), 
where the dark gray region shows the 
embedding of this solution, and 
black dashed and red solid lines 
denote the constant $\eta$ and $t$ slices, respectively. 

It is noteworthy that 
the analytic extension of the solution through $y=0$ 
is an ellipse with the vertices moving at the speed of light. 
As discussed at the end of Sec.~\ref{sec:3A1}, 
the major and minor axes in proper length are 
proportional to $t$ 
and are getting smaller and smaller as $t$ increases from $-\infty$ to $0$.

\medskip
Let us turn our attention to a solution in Region I\hspace{-.1em}I. 
Integrating Eqs.~\eqref{eq:dotChi}--\eqref{eq:dotPhi}, we obtain
\begin{align}
&\sinh \bar{\mathcal{X}}
=\sqrt{\left(1+\bar{l}^{\,2}\right)\tau^2-\frac{1}{1+\bar{l}^{\,2}}}
\\
&\tan \bar{\Phi}=-\frac{l}{1+\bar{l}^{\,2}}\,\tau^{-1},
\end{align}
where, without loss of generality, 
$\tau_0 < \tau<\infty$, in which $\tau_0=1/(1+\bar{l}^{\,2})$, 
and we have chosen $\sinh\bar{\mathcal{X}}(\tau_0)=0$, 
$\tan\bar{\Phi}(\tau_0)=-\bar{l}$,
and the branch $\tan^{-1}\bar{l}<\bar{\Phi}<\pi$, where $\pi/2<\tan^{-1}\bar{l}<\pi$. 
Then, the solution describes an open string on the constant $\bar{\eta}$ slices 
with 
boundaries at spatial infinity. 
This solution in the comoving Cartesian coordinate system takes the form
\begin{align}
\bar{X}(t,y)=-\sqrt{(t/t_0)^4+y^2/\bar{l}^{\,2}},
\end{align}
where $t$ and $y$ have been taken to new parameters. 
Hence, this string on the constant $t$ slices shows 
a hyperbola in the second quadrant on $z=0$ 
and includes an end point at $(x,y,z)=(-(t/t_0)^2,0,0)$ 
moving at the speed of light. 
In Fig.~\ref{fig:1}(e), 
the light gray region shows the world sheet of this solution, 
on which black dashed and blue solid lines denote the constant $\chi$ and $t$ slices, respectively.

The analytic extension of the solution through $y=0$ is a half of hyperbola 
including the vertex with null trajectory.
Note that $\bar{l}$ determines the curvature of the string. For example, 
the string in the limit $\bar{l} \to \infty$ is a straight line, and 
the one with 
$\bar{l}\ll 1$ has high curvature at the vertex.

\subsubsection{$C<1/2$}
\label{sec:3B3}

Let us investigate the solution that has 
constant $\chi$ in the expanding universe with $C<1/2$ (\,i.e., $-5/3<w<-1/3$\,), 
which is filled with dark or phantom energy. 
We call this the self-similar circular string in what follows. 
In the case $l=l_\textrm{c}$, where $l_\textrm{c}$ is defined in Eq.~\eqref{eq:lrange}, 
this is realized with the radius 
$\chi=\chi_\textrm{c}$, 
where
\begin{align}
\chi_\textrm{c}=\tanh^{-1}\frac{1}{\sqrt{2\,(1-C)}}.
\end{align}
Substituting $\chi_\textrm{c}$ and $l_\textrm{c}$ 
into Eq.~\eqref{eq:Phi'}, 
we have the solution $\Phi(\sigma)=\Phi_\mathrm{c}$, where
\begin{align}
\Phi_\mathrm{c}
=\frac{l_\textrm{c}\sigma}{1-2\,C},
\end{align}
where, without loss of generality, we have determined a constant of integration as $\Phi(0)=0$.

The circumferential radius in proper length is
\begin{align}
\label{eq:Lc}
L_\mathrm{c}
=-\frac{C\,t}{\sqrt{2\left(1-C\right)}},  
\end{align}
where $C\,t <0$.  
This depends linearly on $t$,  
as demonstrated at the last of Sec.~\ref{sec:3A1}. For $C<0$ (\,i.e., $-1<w<-1/3$\,), 
$L_\textrm{c}$ increases as the time $t$ proceeds
because of radially outward initial condition. 
For $0<C<1/2$ (\,i.e., $-5/3<w<-1$\,), 
$L_\textrm{c}$ decreases 
as the time $t$ proceeds
because of radially inward initial condition. 
Furthermore, 
$L_\textrm{c}$ satisfies the relation
\begin{align}
L_\textrm{c}H=\sqrt{
\frac{1-C}{2}},
\end{align}
where $H=a^{-1}da/dt=\left(1-C^{-1}\right)t^{-1}$ is the Hubble parameter. 
Hence, the size of the self-similar circular string is at least
larger than a half of the Hubble radius, i.e., $L_\textrm{c}>H^{-1}/2$.

Since $dL_\textrm{c}/dt$ is constant, 
we can interpret that 
the self-similar circular string is 
realized by the balance 
between the string tension and 
the effect of the cosmic accelerated expansion. 
The string tension acts as an attractive force to the self-similar circular string 
and cancels the effect of the cosmic accelerated expansion 
acting as a repulsive force to the string. 
We can find a circular string similar to the self-similar circular string 
in the de Sitter spacetime (\,i.e., $w=-1$\,), 
where there exists no proper homothetic vector field.
Although the circular string is not a self-similar string, 
this keeps its proper circumferential radius constant~\cite{Larsen:1994gh}, which 
is realized by the balance between the 
string tension and the effect of the de Sitter expansion.

In the limit $C \to 1/2$, 
the world sheet of a self-similar circular string approaches the null surface $r=-\lambda$. 
We could found a self-similar circular null string in the case $C=1/2$, 
if we generalized our formalism to the case of a null homothetic vector field. 
For $C>1/2$ (\,i.e., $w<-5/3$\,), there is no circular self-similar string 
as discussed by use of the potential $V$ in Sec.~\ref{sec:3A1}. 
Such a string would be physically forbidden 
because its world sheet would become spacelike.

To conclude whether this model is physically realistic, 
we analyze stability of the self-similar circular string 
under a linear perturbation. 
Let $L(t)$ be the proper radius of the circular string 
that is radially disturbed around $L_\textrm{c}$ given by Eq.~\eqref{eq:Lc} in the form 
\begin{align}
L(t)=L_\textrm{c}\left(1+\delta(t)\right),
\label{eq:perturbation}
\end{align}
where $\delta(t)\ll 1$. 
Substitution of Eq.~\eqref{eq:perturbation} 
into Eq.~\eqref{eq:N-Geq}
yields
the linearized equation in $\delta$
\begin{align}
\ddot{\delta}
+\frac{3\,\dot{\delta}}{t}-\frac{2\left(1-2\,C\right)}{C^2 t^2}\,\delta
=0.
\end{align}
The perturbation 
evolves with $t$ as
\begin{align}
&\delta=\alpha_+\left(t/t_0\right)^{-1+\sqrt{1+2\,\left(1-2\,C\right)/C^2}}
+\alpha_-\left(t/t_0\right)^{-1-\sqrt{1+2\,\left(1-2\,C\right)/C^2}},
\\
&\alpha_\pm=\frac{\delta_0}{2}\pm \frac{\delta_0+\beta_0}{2\sqrt{1+2\,(1-2\,C)/C^2}},  
\end{align}
where $\delta_0=\delta(t_0)$ and $\beta_0=t_0\,\delta(t_0)$. 
The first term is a growing mode for $C<0$ (\,i.e., $-1<w<1/3$\,), 
and the second term is a growing mode for 
$0<C<1/2$ (\,i.e., $-5/3<w<-1$\,). 
Thus, we can conclude that 
a self-similar circular string 
is an unstable equilibrium solution.

\section{summary}
\label{sec:4}

In this paper, 
we have proposed a self-similar string in a 
self-similar spacetime. 
The self-similar string
is defined by the world sheet to which 
a homothetic vector field in a self-similar target spacetime 
is tangent. 
We have investigated the dynamics 
on the basis of 
Nambu-Goto string theory 
and have demonstrated the equation of motion 
to be 
an ordinary differential equation identified with 
the equation of motion in particle mechanics. 
The equation further reduces to a geodesic equation 
in the following cases:  
(i) The homothetic vector field is a Killing vector field (i.e., a cohomogeneity-one string). 
(ii) The homothetic vector field is hypersurface orthogonal. 
(iii) It is the parallel condition [see Eq.~\eqref{eq:cond} for details]. 
Hence, at least in these cases, 
we have obtained the formalism for a self-similar string in a similar manner as 
formulated in the cohomogeneity-one string.

We have applied our formalism to 
a self-similar string in the
Minkowski spacetime or spatially flat FLRW
expanding spacetime with self-similarity.   
In the Minkowski spacetime, 
a self-similar string becomes a straight segment or line, 
which is eventually identified with a cohomogeneity-one string. 
In the expanding spacetime, 
however, 
a self-similar string 
can have nontrivial configuration, 
which is classified 
into two types: 
extended to spacetime boundary and 
confined in a finite region. 
The former includes a straight line solution, 
which also has spatial homogeneity and is linearly stable. 
The latter includes analytically tractable solutions, 
so that we have obtained 
geometrically simple configuration such as an ellipse and a hyperbola 
in the case where $C=1/2$. 
In addition, 
a circular self-similar string for $C<1/2$ 
explicitly provides us 
instructive pictures. 
We have found that 
the solution
is realized by the balance between 
the effect of the cosmic accelerated expansion and the string tension. 
These kinds of solutions evolve linearly in the cosmic time.  
Note that, however, a circular self-similar string 
is an unstable equilibrium solution. 
The result suggests that 
all self-similar string confined in a finite region 
are unstable, and cannot be a candidate for a final state.

We are able to investigate 
a self-similar string 
in these expanding universes further
by using a numerical integration 
and then obtain many nontrivial configurations. 
In addition, it will provide a deeper insight 
to investigate the other choices of a homothetic vector field 
defining a self-similar string (e.g., with twist)
or a target spacetime (e.g., gravitationally collapsing backgrounds). 
Then, we are able to verify the 
validity of Condition (iii).

We can generalize this definition by means of 
the other self-similarity, e.g., kinematic self-similarity. 
It is interesting for future work to 
examine self-similar strings in the other string theories or 
self-similar membranes (which might be a generalization of Ref. \cite{Kozaki:2014aaa}).

\acknowledgments

This work was partially supported by MEXT-Supported Program for the Strategic Research Foundation at Private Universities, 2014-2017~(T.~Igata \& T.~Harada) and 
JSPS KAKENHI Grant Number JP26400282~(T.~Harada) and JP14J01237~(T.~Houri).


\begin{thebibliography}{99}

%
\bibitem{Vilenkin:2000jqa} 
  A.~Vilenkin and E.~P.~S.~Shellard,
  \textit{Cosmic Strings and Other Topological Defects} 
  (Cambridge University Press, Cambridge, England, 2000).  


%
\bibitem{Zwiebach:2004tj} 
  B.~Zwiebach,
  \textit{A First Course in String Theory,}
  (Cambridge University Press, Cambridge, England, 2009).  

%
\bibitem{Ogawa:2008qn} 
  K.~Ogawa, H.~Ishihara, H.~Kozaki, H.~Nakano, and S.~Saito,
  Phys.\ Rev.\ D {\bf 78}, 023525 (2008).
  
%
\bibitem{Burden:1982zb} 
  C.~J.~Burden and L.~J.~Tassie,
  Austral.\ J.\ Phys.\  {\bf 35}, 223 (1982).
  
%
\bibitem{Burden:1984xk} 
  C.~J.~Burden and L.~J.~Tassie,
  Austral.\ J.\ Phys.\  {\bf 37}, 1 (1984).
  
%
\bibitem{Embacher:1992te} 
  F.~Embacher,
  Phys.\ Rev.\ D {\bf 46}, 3659 (1992); 
  Phys.\ Rev.\ D {\bf 47}, 4803 (1993).

%
\bibitem{Kozaki:2009jj} 
  H.~Kozaki, T.~Koike, and H.~Ishihara,
  Classical Quantum Gravity {\bf 27}, 105006 (2010).
  
%
\bibitem{Igata:2012kx} 
  T.~Igata, H.~Ishihara, and K.~Nishiwaki,
  Phys.\ Rev.\ D {\bf 86}, 104020 (2012).

%
\bibitem{Carter:1989bs} 
  B.~Carter and V.~P.~Frolov,
  Classical Quantum Gravity {\bf 6}, 569 (1989).
  
%
\bibitem{deVega:1996mv} 
  H.~J.~de Vega and I.~L.~Egusquiza,
  Phys.\ Rev.\ D {\bf 54}, 7513 (1996).
  
%
\bibitem{Frolov:1996xw} 
  V.~P.~Frolov, S.~Hendy, and J.~P.~De Villiers,
  Classical Quantum Gravity {\bf 14}, 1099 (1997).
  
%
\bibitem{Igata:2009dr} 
  T.~Igata and H.~Ishihara,
  Phys.\ Rev.\ D {\bf 82}, 044014 (2010).
  
%
\bibitem{Igata:2009fd} 
  T.~Igata and H.~Ishihara,
  Phys.\ Rev.\ D {\bf 81}, 044024 (2010).
  
%
\bibitem{Vilenkin:1981kz} 
  A.~Vilenkin,
  Phys.\ Rev.\ D {\bf 24}, 2082 (1981).

%
\bibitem{Gasperini:1990xg} 
  M.~Gasperini, N.~G.~Sanchez, and G.~Veneziano,
  Int.\ J.\ Mod.\ Phys.\ A {\bf 6}, 3853 (1991). 

%
\bibitem{deVega:1993rm} 
  H.~J.~de Vega, A.~L.~Larsen, and N.~G.~Sanchez,
  Nucl.\ Phys.\ B {\bf 427}, 643 (1994).
  
\bibitem{Li:1993qc} 
  X.~Z.~Li and J.~Z.~Zhang,
  Phys.\ Lett.\ B {\bf 312}, 62 (1993). 

%
\bibitem{Larsen:1995vr} 
  A.~L.~Larsen and N.~G.~Sanchez,
  Phys.\ Rev.\ D {\bf 54}, 2483 (1996).

%
\bibitem{Frolov:1988zn} 
  V.~P.~Frolov, V.~Skarzhinsky, A.~Zelnikov, and O.~Heinrich,
  Phys.\ Lett.\ B {\bf 224}, 255 (1989).

%
\bibitem{Ishihara:2005nu} 
  H.~Ishihara and H.~Kozaki,
  Phys.\ Rev.\ D {\bf 72}, 061701 (2005).

%
\bibitem{Ogawa:2008yx} 
  K.~Ogawa, H.~Ishihara, H.~Kozaki, and H.~Nakano,
  Phys.\ Rev.\ D {\bf 79}, 063501 (2009).

%
\bibitem{Harada:2001nh} 
  T.~Harada and H.~Maeda,
  Phys.\ Rev.\ D {\bf 63}, 084022 (2001).

%
\bibitem{Maeda:2001jh} 
  H.~Maeda and T.~Harada,
  Phys.\ Rev.\ D {\bf 64}, 124024 (2001).
  
%
\bibitem{Harada:2001hk} 
  T.~Harada,
  Classical Quantum Gravity {\bf 18}, 4549 (2001).

%
\bibitem{Maeda:2002br} 
  H.~Maeda, T.~Harada, H.~Iguchi, and N.~Okuyama,
  Phys.\ Rev.\ D {\bf 66}, 027501 (2002).

%
\bibitem{Maeda:2002bv} 
  H.~Maeda, T.~Harada, H.~Iguchi, and N.~Okuyama,
  Prog.\ Theor.\ Phys.\  {\bf 108}, 819 (2002).

%
\bibitem{Harada:2002ui} 
  T.~Harada, H.~Maeda, and B.~Semelin,
  Phys.\ Rev.\ D {\bf 67}, 084003 (2003).
  
%
\bibitem{Maeda:2003vc} 
  H.~Maeda, T.~Harada, H.~Iguchi, and N.~Okuyama,
  Prog.\ Theor.\ Phys.\  {\bf 110}, 25 (2003).
  
%
\bibitem{Harada:2003jg} 
  T.~Harada and H.~Maeda,
  Classical Quantum Gravity {\bf 21}, 371 (2004).
  
%
\bibitem{Maeda:2004kw} 
  H.~Maeda and T.~Harada,
  arXiv:gr-qc/0405113.

%
\bibitem{Harada:2006dv} 
  T.~Harada, H.~Maeda, and B.~J.~Carr,
  Phys.\ Rev.\ D {\bf 74}, 024024 (2006).

%
\bibitem{Harada:2007tj} 
  T.~Harada, H.~Maeda, and B.~J.~Carr,
  Phys.\ Rev.\ D {\bf 77}, 024022 (2008).
  
%
\bibitem{Maeda:2007tk} 
  H.~Maeda, T.~Harada, and B.~J.~Carr,
  Phys.\ Rev.\ D {\bf 77}, 024023 (2008).

%
\bibitem{Harada:2008rx} 
  T.~Harada, K.~i.~Nakao, and B.~C.~Nolan,
  Phys.\ Rev.\ D {\bf 80}, 024025 (2009); 
  Phys.\ Rev.\ D {\bf 80}, 109903 (2009).

%
\bibitem{Carr:2005uf} 
  B.~J.~Carr and A.~A.~Coley,
  Gen.\ Relativ.\ Gravit.\  {\bf 37}, 2165 (2005).

%
\bibitem{Schild:1976vq} 
  A.~Schild,
  Phys.\ Rev.\ D {\bf 16}, 1722 (1977).

%
\bibitem{Maartens:1986} 
  R.~Maartens and S.~D.~Maharaj,
  Classical Quantum Gravity {\bf 3}, 1005 (1986). 
 
%
\bibitem{Dabrowski:2003jm} 
  M.~P.~Dabrowski, T.~Stachowiak, and M.~Szydlowski,
  Phys.\ Rev.\ D {\bf 68}, 103519 (2003).
  
%
\bibitem{Larsen:1994gh} 
  A.~L.~Larsen,
  arXiv:hep-th/9408026.

%
\bibitem{Kozaki:2014aaa} 
  H.~Kozaki, T.~Koike, and H.~Ishihara,
  Phys.\ Rev.\ D {\bf 91}, 025007 (2015).

\end{thebibliography}
\end{document}